# Peering into the Dark (Ages) with Low-Frequency Space Interferometers

## Using the 21-cm signal of Neutral Hydrogen from the Infant Universe to probe Fundamental (Astro)Physics

**Contact Scientist: Prof. dr. L.V.E. (Léon) Koopmans[1]**

(Kapteyn Astronomical Institute, University of Groningen, the Netherlands)

**Core Proposing Team:**

Rennan Barkana (Tel-Aviv), Mark Bentum (TUE), Gianni Bernardi (Bologna/SKAO-SA), Albert-Jan Boonstra (ASTRON), Judd Bowman (ASU), Jack Burns, (Colorado/Boulder), Xuelei Chen (NOAC), Abhirup Datta (IIT Indore), Heino Falcke (Radboud), Anastasia Fialkov (Sussex), Bharat Gehlot (ASU), Leonid Gurvits (JIVE/TUD), Vibor Jelić (IRB), Marc Klein-Wolt (Radboud), Léon Koopmans (Kapteyn), Joseph Lazio (JPL, CIT), Daan Meerburg (VSI, Groningen), Garrelt Mellema (Stockholm), Florent Mertens (Kapteyn, Groningen), Andrei Mesinger (SNS), André Offringa (ASTRON), Jonathan Pritchard (Imperial College), Benoit Semelin (Obs. de Paris), Ravi Subrahmanyan (RRI), Joseph Silk (Oxford), Cathryn Trott (Curtin), Harish Vedantham (ASTRON), Licia Verde (ICC), Saleem Zaroubi (Kapteyn/Open Univ.), Philippe Zarka (Obs. de Paris)

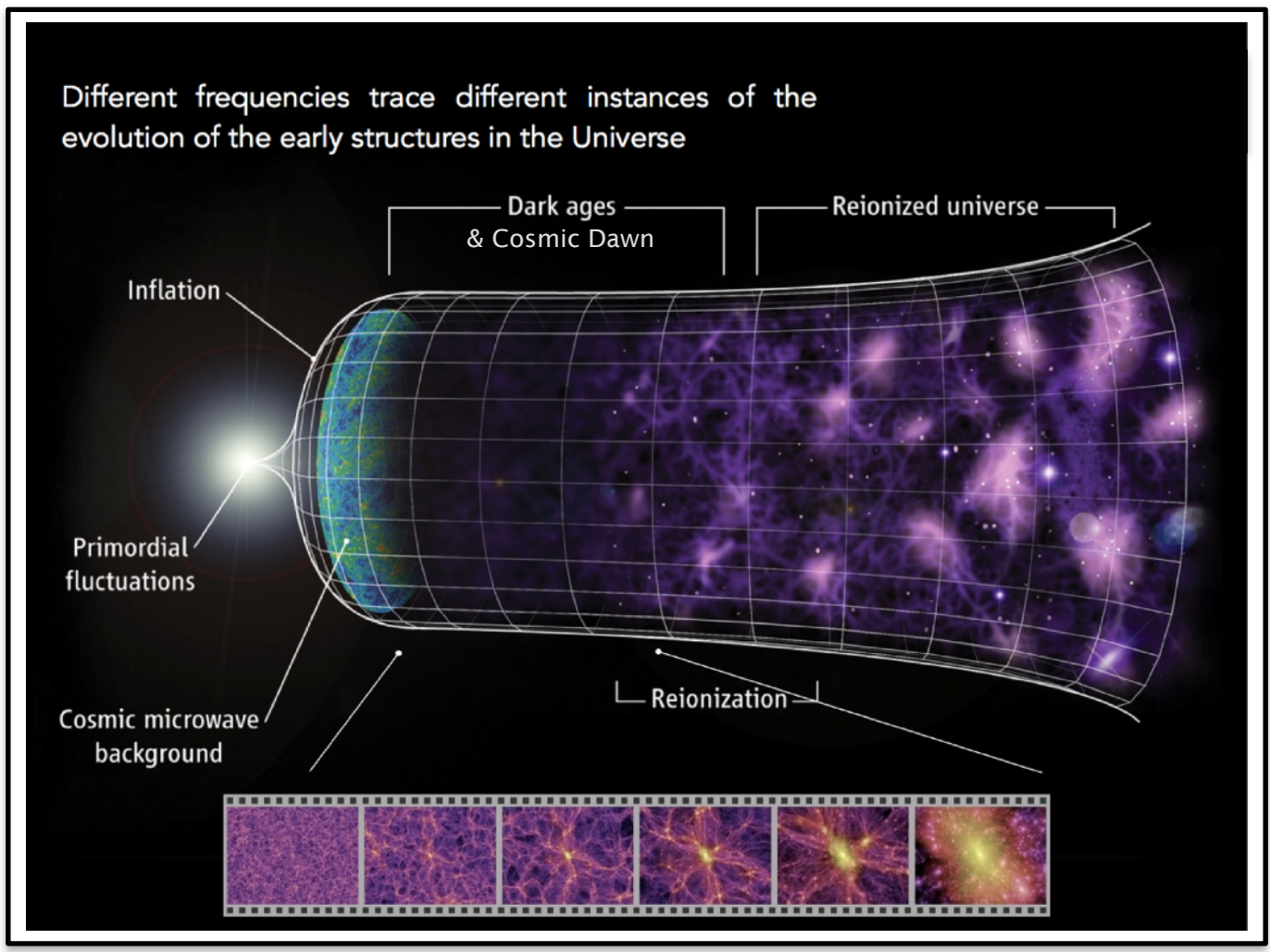

**ESA Voyage 2050 — White Paper**

[1] koopmans@astro.rug.nl; Kapteyn Astronomical Institute, University of Groningen, P.O.Box 800, 9700AV, Groningen, Netherlands; Phone: +31(0)503636519

# 1. Executive summary

The first billion years of the Universe (its 'infancy') is one of the least explored and understood eras in the Universe, despite enormous observational progress in recent years with optical/uv, infrared and sub-mm ground- and space-based instruments. However, between the first image of the Universe from the Cosmic Microwave Background (CMB) and the first stars and galaxies becoming visible, some half to one a billion years later to large ground- and space-based observatories, observational probes remain incredibly scarce. *Redshifts between z~11-1100 are currently devoid of any direct observational probes.* Even JSWT in the 2020s will only be able to push direct observations a little further in redshift (z~15) and only for a limited number of bright sources and over a limited field of view.

*Neutral hydrogen, on the other hand, pervades the infant Universe and its redshifted 21-cm signal allows one to chart the Universe from ~10 kpc to ~100 Mpc scales*, via absorption and emission against the CMB and, potentially, via high-z radio sources. This signal allows one to probe astrophysical processes such as the formation of the first stars, galaxies, (super)massive black holes and enrichment of the pristine gas from z~6 to z~30, as well as fundamental physics related to gravity, dark matter, dark energy and particle physics at redshift beyond that. *As one enters the so-called Dark Ages (z>30), the Universe becomes pristine. Density perturbations are linear, and its physics is thought to be well understood. Deviations from linear-theory predictions of the 21-cm signal would thus usher new physics that nearly impossible to probe at lower redshift where signals mix with those from complex astrophysical processes.*

Ground-based low-frequency radio telescopes — e.g. LOFAR, MWA, OVRO-LWA and NenuFar and the planned HERA and SKA — aim to detect the spatial fluctuations of the 21-cm signal. Complementary, global 21-cm experiments — e.g. EDGES [recently claiming a 21-cm signal detection at z~17], and SARAS — aim to measure the sky-averaged 21-cm signal. All these instruments focus on redshifts z~6 to ~25 though. Pushing these observations to higher redshifts from the ground is impossible due to radio-frequency interference (RFI) and the ionosphere, even with the most powerful and isolated next-generation instruments such as SKA and HERA. *New space-based radio-interferometers are critical in 2030+ to push observations of the spatially-varying 21-cm signal beyond z~25.*

Escaping RFI and the ionosphere has motivated space-based missions, such as the Dutch-Chinese NCLE instrument (currently in lunar L2), part of the Chinese Chang-e'4 mission, the proposed and partially funded US-driven lunar or space-based instruments DAPPER and FareSide, the lunar-orbit interferometer DSL (China), and PRATUSH (India), all missions to probe the global average or spatial fluctuations (FARSIDE, DSL) of the 21-cm signal. *ESA is currently not involved in any of these pioneering missions, despite the tremendous discovery space that it will open and the promise for uncovering new (astro)physics.* The importance of this era is emphasised by the numerous related white papers submitted to the 2020 US Decadal Review, involving many researchers from the global community.

> Some **Key Questions** that low-frequency space-based radio interferometers — sensitive enough to probe the 21-cm signal from the **Dark Ages (z>30)** — will be able to probe, are related to: (i) inflation, (ii) the nature of dark matter and dark energy, (iii) gravity on scales currently inaccessible, (iv) particle physics (e.g. WIMPS, neutrinos, axions), (v) gravitational waves, (vi) non-gaussianity and (vii) primordial black holes. The same 21-cm signal observations at somewhat lower redshifts (z~30-10) can also shed light, via tomography (impossible even with SKA or HERA), on never before seen astrophysical processes during the **Cosmic Dawn**, such as signatures of the formation of the first star, galaxies, stellar-remnant black holes and the synthesis of elements heavier than helium.

To push beyond the current z~25 frontier, though, and measure both the global and spatial fluctuations (power-spectra/tomography) of the 21-cm signal, low-frequency (1-100MHz; BW~50MHz; z>13) space-based interferometers with vast scalable collecting areas (1-10-100 km²), large filling factors (~1) and large fields-of-view (4π sr.) are needed over a mission lifetime of >5 years. *In this ESA White Paper, we argue for the development of new technologies enabling interferometers to be deployed, in space (e.g. Earth-Sun L2) or in the lunar vicinity (e.g. surface, orbit or Earth-Moon L2), to target this 21-cm signal.* This places them in a stable environment beyond the reach of most RFI from Earth and its ionospheric corruptions, enabling them to probe the Dark Ages as well as the Cosmic Dawn, and allowing one to investigate new (astro)physics that is inaccessible in any other way in the coming decades.

## 2. The first billion years of the universe

Over the last hundred years, astronomers have discovered that we live in an ever-expanding Universe which started 13.8 billion years ago in a very hot and dense state, commonly known as the Big Bang. The model which provides an excellent description of both the earliest phases and the more recent expansion is based on General Relativity and postulates that apart from ordinary, baryonic, matter, the Universe also contains a substantial amount Dark Matter and Dark Energy. This model is known as *'Λ-Cold Dark Matter (ΛCDM)'* because of the specific properties of the Dark Matter (cold) and the Dark Energy (a constant Λ). In this model, the first hundred millions of years are a period in which the Universe, due to its expansion, became globally more and more rarefied and cold. At the age of about 380,000 years, the temperature dropped to about 3000 K and collisions were no longer be able to keep the electrons from combining with the (mostly hydrogen) nuclei. As the Universe became neutral, baryonic matter decoupled from radiation, leaving the latter free to travel and allowing the former to follow the gravitational pull of tiny dark matter concentrations (Pritchard & Loeb 2012). Over time, these local concentrations of dark and baryonic matter grew large enough to allow the formation of the first stars, called the *Cosmic Dawn* (z~10-30). The era after decoupling and before the first stars (z>30) formed is commonly known as the *Dark Ages* and is the main focus of this White Paper.

**Probing the Dark Ages (and beyond) via the redshifted 21-cm signal of neutral hydrogen**

The 21-cm hyperfine line of neutral hydrogen — central to the science in this white paper — arises from the magnetic interaction between the proton and electron spins in neutral hydrogen. A spin-flip from the state with spins aligned to spins anti-aligned will emit a photon with a wavelength of 21.1 cm or a frequency of 1420 MHz. The connection of the atomic physics to astrophysics can be quite subtle and is an active area of research (see Furlanetto et al. 2006, Pritchard & Loeb 2012, for reviews). For our purpose, we care about the emission or absorption of 21-cm wavelength light from hydrogen gas seen against some background source at some point in the cosmic time (called the '21-cm signal' hereafter). The intensity of the 21-cm signal depends upon the degree of ionisation, density, temperature, and excitation state of the hydrogen gas in the cloud. Following the convention of radio astronomers, who describe the observed intensity of light in terms of a brightness temperature ($I_\nu = 2kT_b \nu^2/c^2$), one can write the differential brightness temperature for the 21-cm signal seen against some backlight as (Pritchard & Loeb 2012):

$$\delta T_b \propto \rho_{HI} \left( \frac{T_S - T_{back}}{T_S} \right) (\partial_r v_r)^{-1}. \qquad \text{(Eqn. 1)}$$

Here $\rho_{HI}$ is the density of neutral hydrogen and $T_{back}$ describes the intensity of background radiation. $\partial_r v_r$ is the local radial velocity gradient that controls the mapping between local and observed wavelengths – typically this will be determined by the cosmic expansion with minor corrections due to local peculiar velocities. $T_S$ is a spin temperature that describes the ratio of atoms in the spin-aligned versus spin-anti aligned states ($n_{\uparrow\uparrow}/n_{\uparrow\downarrow}=3\exp[-h\nu_{21}/kT_S]$). The spin temperature is a key factor in interpreting the 21-cm signal (Barkana & Loeb 2005, Pritchard & Furlanetto 2007) and varies between the temperature of the CMB and the gas. At the earliest times (e.g. during the Dark Ages), it is set through thermal collisions between neutral hydrogen and other particles, which locks $T_S=T_{gas}$ (Allison & Dalgarno 1969; Zygelman 2005). As cosmic expansion dilutes the number density of hydrogen atoms, collisions become rare and the scattering of CMB photons relaxes $T_S=T_{CMB}$. Finally, as stars begin to illuminate the universe (during the Cosmic Dawn) with their light, a process of resonant scattering of Lyman alpha photons, known as the Wouthuysen-Field effect, drives $T_S=T_{gas}$ (Wouthuysen 1952, Field 1959). Since the gas temperature varies significantly over cosmic time, the 21-cm signal can be viewed as a thermometer to measure the impact of cooling due to cosmic expansion or heating due to X-rays from stellar sources (Pritchard & Furlanetto 2007) or exotic physics, e.g. dark matter decay (Furlanetto et al. 2006; see also below). We typically look for the 21-cm signal as a spectral distortion of emission or absorption against some background radio source. For a very bright radio source, such as a radio quasar, the 21-cm signal will always be seen as a series of absorption lines – the 21-cm forest (Furlanetto 2006). Most generally for mapping the Cosmic Dawn or the Epoch of Reionization (EoR; see Fig.1), where the CMB itself is the backlight, the 21-cm line can be seen either in absorption (when $T_S<T_{CMB}$) or emission ($T_S>T_{CMB}$) depending upon the gas temperature and the degree of coupling. The 21-cm signal is therefore extremely rich in information about the early universe and its interpretation depends upon our ability to model the physical properties of hydrogen gas and the impact of radiation – Lyman alpha, ionising or heating – from the first luminous sources (Pritchard & Loeb 2008,

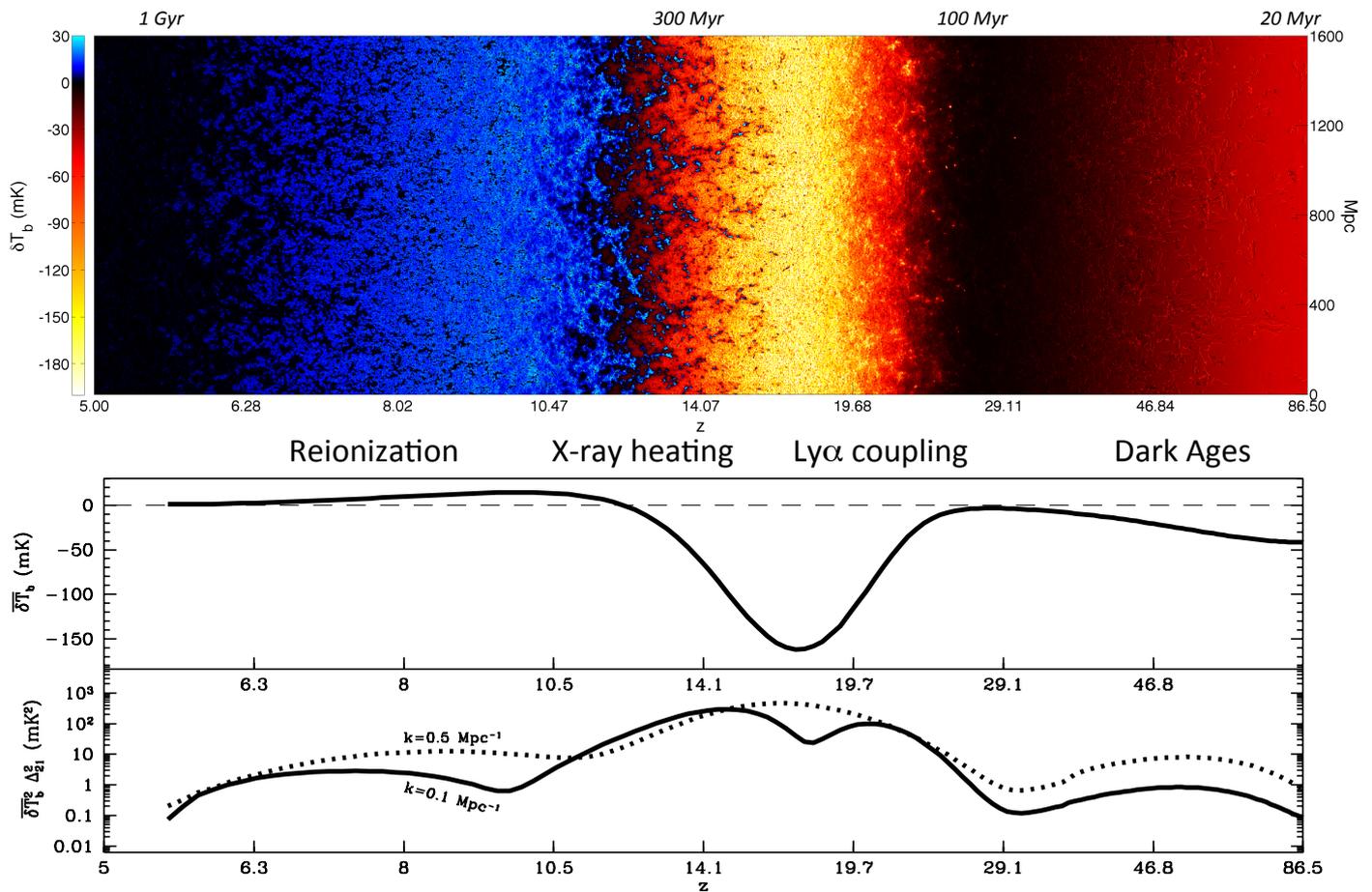

Fig.1: 2D slice through a mock 21-cm lightcone (top), corresponding global global signal evolution (middle), and the 3D power spectra at two wave numbers (bottom). Some major cosmic milestones can be seen from right to left: (i) Wouthuysen-Field (or Lyman-α) coupling (Wouthuysen 1952, Field 1959), when Lyman-band photons efficiently couple the spin temperature to the gas temperature (black to yellow); (ii) epoch of heating, when galactic X-rays with long mean free paths heat the predominantly neutral IGM (yellow to blue); (iii) epoch of Reionization, when ionising photons drive the final major phase transition of our Universe (blue to black). The timing and the patterns of the signal encode a wealth of information about the first galaxies. This figure corresponds to the fiducial model of Mesinger et al. (2016), though recent studies calibrated to high-z galaxy observations suggest that the Cosmic Dawn epochs of WF coupling and X-ray heating might occur later, with significant overlap with the EoR, due to less efficient star formation (e.g. Park et al. 2019).

Mesinger et al. 2011) during the Cosmic Dawn and after. *Before the first sources emit radiation, however, astrophysics was physics. The Dark Ages are the main focus of this white paper.*

## 3. Cosmic eras in the infant universe: from physics to (g)astrophysics

Below, we focus on the Dark Ages and early Cosmic Dawn — which are both extremely hard to observe from earth — and on the (astro)physical processes that can be probed using the 21-cm signal. We do not discuss the Epoch of Reionization (Fig.1; z<10) further, since it is expected to be fully characterised in the 2020s by current and forthcoming ground-based instruments, and does not require space-based missions.

**(1) Dark Ages:** *The 21-cm signal of neutral hydrogen from the Dark Ages (DA) — a period of cosmic history preceding star formation during the Cosmic Dawn — is predicted to be a unique probe of cosmology and fundamental physics.* A measurement of the 21-cm signal from this era is expected to reveal unprecedented details about the early Universe, such as precision measurements of the scale dependence of primordial fluctuations (e.g. Meerburg et al, 2016), their possible deviations from Gaussian statistics (e.g. Meerburg et al, 2017) and even probe primordial gravitational waves through tidal effects and curl lensing (Schmidt et al, 2014, Book et al, 2011, Hirata et al, 2018, Masui et al, 2017, Sheere et al, 2017 and Ansari et al, 2018). Furthermore, the Dark Ages signal will allow testing of fundamental theories in a completely unexplored regime. The shape of this signal depends on the underlying cosmological model (expansion history), nature of dark matter and the radio background radiation. Assuming the standard cold dark matter cosmology, conventional expansion history of the Universe (e.g., Planck 2018) and the CMB as a background light, the 21-cm signal can be calculated precisely. *Any deviations from the predicted signal could be a manifestation of exotic physics.* For instance, extra radio background light in addition to the CMB would completely alter the shape of both the sky-averaged (global) signal and the power spectrum in a model-dependent way (e.g., Fialkov & Barkana, 2019). Such deviations could be contributed to by neutrinos (Chianese et al., 2019), dark matter decay (Fraser et al., 2018, Pospelov et al., 2018), or by superconducting strings (Brandenberger et al., 2019). The Dark Ages 21-cm signal will also be an important probe of dark matter physics (e.g., Slatyer 2013; Slatyer 2016a,b; Tashiro et al. 2014) allowing one to test different particle physics models of dark matter in currently unconstrained regimes. For example, the annihilation rate of Weakly Interacting Massive Particles (WIMPs) is expected to be higher in the dense environment of the early Universe than today (e.g. Crelli et al. 2009). By-products of dark matter annihilation (or decay) will heat and ionise the gas imprinting characteristic signature in the 21-cm signal. In another class of models where dark matter directly interacts with baryons (e.g., Tashiro et al. 2014), thermal history of the gas is also modified. A smoking gun signature of interacting dark matter is in the 21-cm power spectrum where, if the interaction cross-section is velocity-dependent, enhanced Baryon-Acoustic oscillations are expected (e.g., Fialkov et al. 2018, Munoz et al. 2018). Finally, models in which dark matter is warm or have a large coherence scale (e.g., ultra-light axions, sterile neutrino) predict delayed star formation and leads to an extended Dark Ages.

**(2) Cosmic Dawn:** *The Cosmic Dawn (CD) of the first galaxies heralded the end of the Dark Ages and the birth of structure.* Baryons cooled onto dark matter halos, collapsing to form the first stars, black holes and galaxies. Their light spread throughout the Universe, eventually heating and ionising virtually every baryon. The CD culminated with the Epoch of Reionization (EoR), whose timing we have recently began to nail down thanks to observations of the CMB (e.g. Planck collaboration 2016), high-redshift QSOs (e.g. Mortlock et al. 2011, Banados et al. 2018) and galaxies (e.g. Mason et al. 2019). However, the CD should contain much more information than just the EoR. Ionising photons responsible for the EoR have relatively short mean free paths through the IGM (of order kpc). However, soft UV and X-ray photons from the first galaxies can travel distances of ~100 Mpc. These radiation fields permeate the Universe while the Cosmic HII regions from the EoR are still proximate to the nascent galaxies. Empirical scaling relations of the soft UV and X-ray luminosities to star formation rates (SFRs) of local galaxies (e.g. Mineo et al. 2012, Brorby et al. 2016, Lehmer et al. 2016) suggest that the soft UV and X-ray backgrounds established during the CD are sufficiently strong to couple the spin temperature of the cosmic gas to its kinetic temperature (allowing us to see the cold gas in contrast against the CMB), and subsequently heat the gas to temperatures above the CMB (McQuinn & O'Leary 2012, Pacucci et al. 2014; see Fig.1). *Indeed the CD is expected to result in the strongest 21-cm signal*, with the adiabatically-cooling IGM providing a large contrast to the hot IGM proximate to the first X-ray sources (Fig.1). What can we learn from CD observations? The timing and spatial structure of the cosmic 21-cm signal is sensitive to the nature of the sources. The very first galaxies which start the CD are likely hosted by so-called mini-halos (with masses of $10^{6-8}$ $M_{sun}$) accreting their gas through $H_2$ cooling (e.g. Haiman et al. 1996). Mini-halos likely have very different ISM properties and star formation efficiencies compared to later galaxies (e.g. Xu et al. 2016, Koh & Wise 2018). Due to their relatively pristine gas reservoir, they likely host stars which were more massive and brighter, so-called Population III stars (Tumlinson & Shull 2000, Schaerer 2002, Yoshida et al. 2008). Although mini-halo galaxies likely start the CD, star formation inside them is soon sterilised with the buildup of an $H_2$ dissociative background (e.g. Holtzenbauer & Furlanetto 2012, Fialkov et al. 2013). Atomic hydrogen subsequently provided the dominant cooling mechanism for galaxies inside $>\sim 10^8$ $M_{sun}$ halos. Such dwarf galaxies are still far too faint to be directly observed. However, we hope to be able to extrapolate some of their properties from the brighter high-z galaxies to be observed with JWST (e.g. Senchya et al. 2019). These first generations of galaxies host massive stars, whose soft UV radiation couples the spin temperature to the gas temperature, allowing us to see the cold IGM against the CMB. The timing of this WF coupling tells us when these first stars appeared and disappeared, what was their typical IMF, and how efficient was star formation. The stars in the first galaxies also are responsible for heating the IGM, likely through high-mass X-ray binaries (HMXBs) and the hot ISM created following supernovae events (e.g. McQuinn 2012). The X-rays from HMXBs and the hot ISM can permeate deep into the neutral IGM, well before the bulk of the EoR, partially ionising it and heating it. The spatial structure of this epoch of heating (EoH) is also a sensitive probe of the nature of the dominant X-ray sources (e.g. Pacucci et al. 2014, Fialkov et al. 2015), as well as the ISM properties of the host galaxies which attenuate the emerging soft X-rays (e.g. Das et al. 2017). Similarly to the DA, the CD could also hold clues about even more exotic radiation sources. If dark matter annihilation is an efficient process, the resulting shower could provide a fairly uniform background of X-rays, whose spatial signature is dramatically different to that of astrophysical sources (e.g. Valdes et al. 2013, Evil et al. 2014, Honorez et al. 2016). Moreover, if the recent, putative EDGES detection of an absorption signature at z~17 is proven to be cosmological (Bowman et al. 2018); see however Hills et al. 2018, Draine & Miralada-Escudé 2018, Bradley et al. 2019), the spatial structure of this epoch could tell us either about a population of radio loud mini QSOs (Ewall-Wice et al.

2018) or an exotic interaction between DM and baryons (Mūnoz & Loeb 2018, Barkana 2018). Indeed, the later mechanism could allow us to more easily measure baryonic acoustic oscillations at redshifts of z~20 (McQuinn & O'Leary 2012, Munoz 2019). In summary, the timing and structure of the Cosmic Dawn 21-cm signal will result in an invaluable physical bounty. *By mapping the Cosmic Dawn, we will be able to understand the nature of the first stars, galaxies, (super-massive) black holes, the stellar and ISM properties, and possibly even dark matter.*

**Global Signal versus Spatial Intensity Fluctuations;** Complementary aspects of the 21-cm signal can be studied via two different approaches. The sky-averaged or global signal (monopole in CMB parlance) can be measured by a spectrometer connected to a single dipole antenna (see Fig.1, bottom panel and Sections 5 & 6). Alternatively, angular fluctuations in the 21-cm brightness can be studied at different frequencies via interferometric techniques (see Fig.1, top panel and Sections 5 & 6). The global and spatially fluctuating signals are analogous to the diffuse 2.7K blackbody and anisotropies seen in the CMB respectively. Unlike the CMB, the 21-cm signal, being a spectral line, allows for line-of-sight tomography of the IGM. The measured 21-cm differential brightness temperature (Eqn.1) primarily depends on the radiation temperature, gas temperature and gas ionisation fraction (assumed zero during the CD/DA). In the absence of stellar sources of radiation in the dark ages, all three quantities are strongly constrained by standard baryonic physics applied to an adiabatically expanding Universe permeated by CMB photons. Any departure from theoretical expectations of the global radio spectrum in the Dark Ages (Fig.1) requires an exotic mechanism to substantially alter the physical condition of baryonic matter, and/or the cosmic radiation background on cosmic scales (see above). An accurate global radio spectrum measured by a single dipole (Section 5 & 6) can therefore reveal the presence of a new mechanism to deposit energy into gas on cosmological scales. The predictions are similarly strong for the spatial fluctuations of the 21-cm signal from the Dark Ages, which in the absence of luminous sources (no non-linear structure formation and baryonic feedback), are largely governed by linear structure formation. The resulting baryonic density fluctuations and peculiar velocity have been computed for the standard CDM paradigm (Section 7) and yield relatively stringent predictions for the power spectrum of the 21-cm signal that interferometry can measure. Measured departures will therefore significantly evolve our understanding of the nature of dark matter and its influence on structure formation. We discuss both approaches in Section 6 and 7, although the focus in Section 7 will be on future (in 2030+) interferometric 21-cm signal observations of the Dark Ages and early Cosmic Dawn.

## 4. General motivation and selected key science drivers for 21-cm signal observations

*The Comic eras discussed above constitute the transitional steps that take our Universe from its early linear phases to the highly non-linear Universe that we currently see around us.* Each of these eras ushers the Universe into a new realm, and hence probes different physical processes that can be used to understand the onset of structure formation in the Universe as well as the very nature of the initial conditions in the Universe. The redshifted 21-cm line is uniquely suited to study these phases of the Universe's evolution for a number of reasons. Firstly, it probes the bulk of the hydrogen in the IGM and provides us with a global picture of physical state of the Universe. Secondly, since it comes from a specific wavelength, it allows a detailed tomographic reconstruction of the Universe's history as a function of redshift. Thirdly, it is one of the very few observables (if not the only one) that are accessible to us in this redshift range, either from the ground or, for the longer wavelengths, from space. During the Dark Ages of the Universe the 21-cm line allows us to detect the intergalactic medium in absorption against the CMB (Hogan & Rees 1979; Scott and Rees 1990; Madau, Meiksin & Rees 1997). Since in this stage the matter fluctuations in the Universe are still mostly well within the linear regime of gravitational instability, they faithfully reflect the initial conditions of the Universe in a very detailed manner. Hence, the redshifted 21 cm can be used to e.g. constrain models of inflation or its

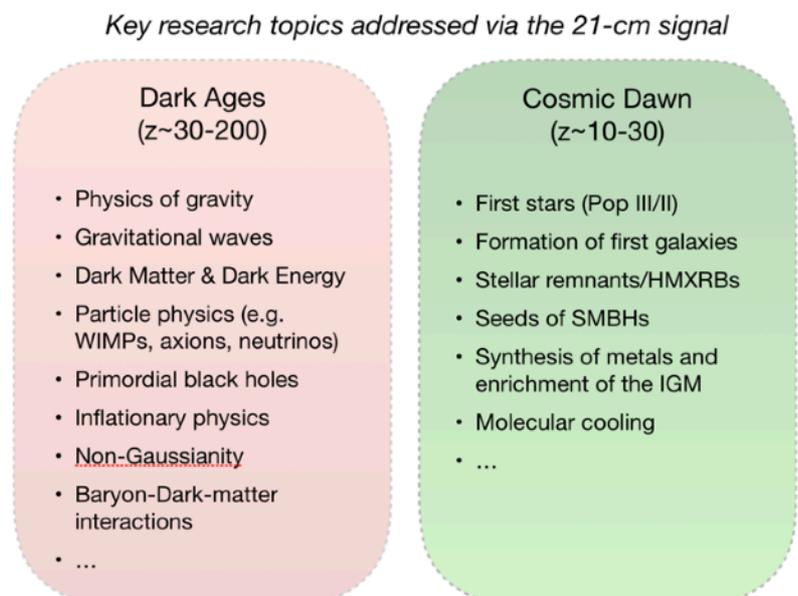

Fig.2: Some of the key questions that the 21-cm signal from the Dark Ages and the Cosmic Dawn can address.

alternatives, through the study of non-Gaussianity with high order statistics (see Section 3 for more details and references). Such data promises to be even more constraining than the CMB since it covers a wide range of redshifts. The 21-cm line from this era can also be used to study many other topics, e.g., extreme density peaks (as progenitors of super massive black holes), annihilating and decaying dark matter particle candidates, etc. During the Cosmic dawn phase the first objects are expected to form and allow us to study the interplay between the Lyman-alpha coupling to the cooler baryons and of the spin temperature (Wouthuysen 1952; Field 1958; 1959) and x-ray heating of the IGM (Maudau et al 1997; Pritchard & Loeb 2010). During this era the IGM is still mostly neutral but the first stars and X-ray sources leave a marked imprint on the fluctuations of the spin temperature in the IGM, most significantly on the contrast between absorption and emission region of the 21-cm signal. The structure and evolution of this spectrum of negative and positive fluctuation shed light on the processes that formed the first objects, when they formed, and how. This is also the period in which the supermassive black holes that we detect almost routinely now between redshift 6 and 7 (Fan et al. 2003; 2006; Mortlock et al., 2011). Therefore exploring this era might answer the puzzle of super massive black hole formation and the rapid build up of their mass in such a short period in the early history of the Universe. This period might also reveal to us new physics as the result from the EDGES telescope tentatively suggest (Bowman 2018). Of course, also here there is a great potential to address cosmological questions. The last of these phases is the Epoch of Reionization (Figure 1), which is targeted by many current telescopes. The 21-cm signal from this epoch informs us about the onset of the Reionization process, and how long it took to permeate the whole intergalactic medium to fill the Universe (see e.g., Zaroubi 2013). *The answer to the questions that arise in the context of studying the Dark Ages, Cosmic Dawn (and the Epoch of Reionization) relate to many fundamental issues in cosmology, galaxy and quasar formation, and the physics of very metal poor stars; all at the heart of many research issues in modern astrophysics.*

**Some Key Science Drivers to observe the Dark Ages:**

Tomography of the 21-cm signal at observed frequencies of ~1 to ~100 MHz thus provides access to a previously unaccessible volume of parameter space and samples a very large number of independent modes describing the primordial fluctuation field. This will open a new window on information content about the properties of the early universe, and is *highly complementary to that embedded in the CMB temperature and polarisation signals.*

- **$f_{NL}$, single vs multi-field inflation:** Because of the stringent Planck limits in the r-$n_s$ plane, viable inflationary models today mostly have very low r. This means that CMB polarisation experiments need exquisite sensitivity and wide sky coverage to qualitatively improve our understanding of inflation. However there is no guaranteed prediction of r. Primordial non-gaussianity provides a robust but challenging complementary probe of inflation, even the simplest inflationary models predict small but non-zero deviations from gaussianity, while most multi-field inflation models are expected to have $f_{NL} \sim 1$ and many predict features in $f_{NL}$. The errors bars on $f_{NL}$ scale with the (inverse of the) number of independent modes. The limited number of modes in the CMB (~$10^6$) and in large-scale galaxy surveys (~$10^8$) strongly motivates exploration of 21-cm Dark Ages cosmology at z~25-75 (20-60 MHz where up to $10^{12}$ modes can be explored. Attainment of a robust limit on (or detection of) primordial non-gaussianity would provide the ultimate probe of generic inflation.

- **Primordial black holes:** Primordial black holes are the only dark-matter candidate that avoids introducing new beyond-standard-model physics and are motivated by gravitational wave observations, but the remaining window for PBH to be a significant DM contributor is limited to the sub-lunar and to the solar-mass range. One needs boosted scale-dependent and possibly non-Gaussian conditions, available in many and possibly all inflationary models. These may lead to spectral power boosts, once accretion and Poisson clustering are included, as well as to IGM heating, that can be uniquely probed via very low frequency 21-cm Dark Ages observations, during an era well before any astrophysical sources have formed that complicate interpretations.

- **Dark-matter annihilation and decay:** Identification of a weakly interacting dark-matter particle candidate is the holy grail of particle astrophysics. The Dark Ages offer a unique forum for exploring long-lived particle annihilations or decays because the 21-cm signal and especially its fluctuations are highly sensitive to electromagnetic energy injection. This allows study of dark-matter models in a complementary way to direct detection, via precision spin temperature measurements, should provide order-of-magnitude improvements over the best current Dark-Ages limits from CMB constraints. The advantage of measuring DM annihilations and decays during the dark ages is that it provides a direct probe of exotic physics, uncontaminated by astrophysical processes.

- **Ultra-light Axions:** The axion is a hypothetical dark-matter particle that would explain the lack of charge-parity violation in strong interactions, and that naturally emerges from string theory. It represents an interesting dark-matter candidate and many experimental searches are being planned. Moreover, the ultralight axion uniquely combines quantum and astronomical scales, with its de Broglie wavelength potentially measurable via the ultra-light axions mass, and potentially mitigates small-scale observational tensions within the ΛCDM paradigm. Ultra-light axion signatures on the matter power spectrum appear on very small scales, making observations of 21-cm absorption against the CMB during the Dark Ages a unique observable to set competitive constraints on its mass, complementing and potentially surpassing CMB probes.

- **SMBH:** There is still no convincing scenario for the origin of the seeds of supermassive black holes that are located in the centres of most massive galaxies. If these are seeded by a population of primordial black holes, gas accretion around PBHs during the cosmic dark ages would leave a unique signature in the 21-cm signal. The presence of a population of primordial black holes would change the standard (ΛCDM-based) interpretation of the CMB anisotropy signal, opening up degeneracies among cosmological parameters and relaxing current ΛCDM-based constraints, hence providing complementarity of the two probes.

- **DM-baryon relative velocity:** Baryons and Cold Dark Matter have supersonic relative velocities after recombination. This relative velocity suppresses the growth of matter fluctuations, and the amplitude of the small-scale power spectrum is modulated on scales over which the relative velocity varies, and leaves an imprint in the small-scale, very high-z matter power spectrum. DM models that suppress the small scale spectrum (such as warm dark matter and ULAs) would cancel this modulation. Therefore, this observable can both test ΛCDM predictions and DM models.

## 5.   Current/Planned Ground-based observatories

A wide range of low-frequency arrays have been either used, or specifically designed, to detect the redshifted 21-cm signal of neutral hydrogen over the past decade (Fig.3). Some have already been decommissioned, or are being upgraded. Among them are the 21CMA (China), GMRT (India) and PAPER (US/South Africa), the latter no longer active, an LOFAR (Netherlands), MWA (Australia), OVRO-LWA (US) instruments, and planned arrays such as HERA (SA) and the SKA (AU/SA). All instruments have the Epoch of Reionization as their main target, since the signal-to-noise is expected to be largest, and least affected by the ionosphere and (extra)Galactic foregrounds, and RFI. Some target the Cosmic Dawn as well, but detections are only feasible in case of very large ('exotic' 21-cm signal; see above). Below a short summary is provided for each of the these instruments, each with their own specific design and technology, allowing one to set the scene for future space-space instruments.

**21CMA[2]:** The *21 Centi-Meter Array* (e.g. Zheng et al. 2016) is situated in the Tianshan Mountains of western China. It is a meter-wave interferometric array designed to probe the 21-cm signal from the Cosmic Dawn and the Epoch of Reionization at z=6-27. The array was constructed between August 2005 and July 2006 and upgraded by July 2010. It consists of 81 pods, each with 127 log-period antennas, which are deployed along two perpendicular arms of 6 and 4 km, respectively. A field of 10-100 square degrees, centred on the North Celestial Pole (NCP), is imaged 24 hours per day in the frequency range of 50-200 MHz and with a spectral resolution of 24 kHz. Coherent uv-data at each frequency channel are being accumulated to meet the desired sensitivity of a statistical detection of the 21-cm signal, and advanced RFI and foreground removal techniques have been developed.

**GMRT[3]:** The *Giant Metrewave Radio Telescope* is a radio interferometer consisting of 30 antennas, each with a diameter of 45m. Fourteen of these are arranged in a 1-km dense central core which allows high-brightness sensitivity required to search for the 21-cm signal (Pen et al. 2009). The longest separation between antennas is about 25 km. Data were taken over 5 nights in December 2007, accumulating about 40h. The observations were centred on PSR B0823+26, used to calibrate the instrument and ionosphere. The primary beam has a full width half-maximum (FWHM) of 3.1◦ and a maximum angular resolution of about 20 arc-sec. The bandwidth covers a frequency range from 139.3 to 156.0 MHz in 64 frequency bins of 0.25 MHz each with a time resolution of 64 s. This corresponds to a redshift range of z= 8.1–9.2, providing an upper limit of (248 mK)^2 for k=0.50 h Mpc$^{-1}$ at z=8.6 at the 2-sigma level (Pagica et al. 2013).

---

[2] http://english.nao.cas.cn/Research2015/rp2015/201701/t20170120_173603.html

[3] http://www.gmrt.ncra.tifr.res.in/

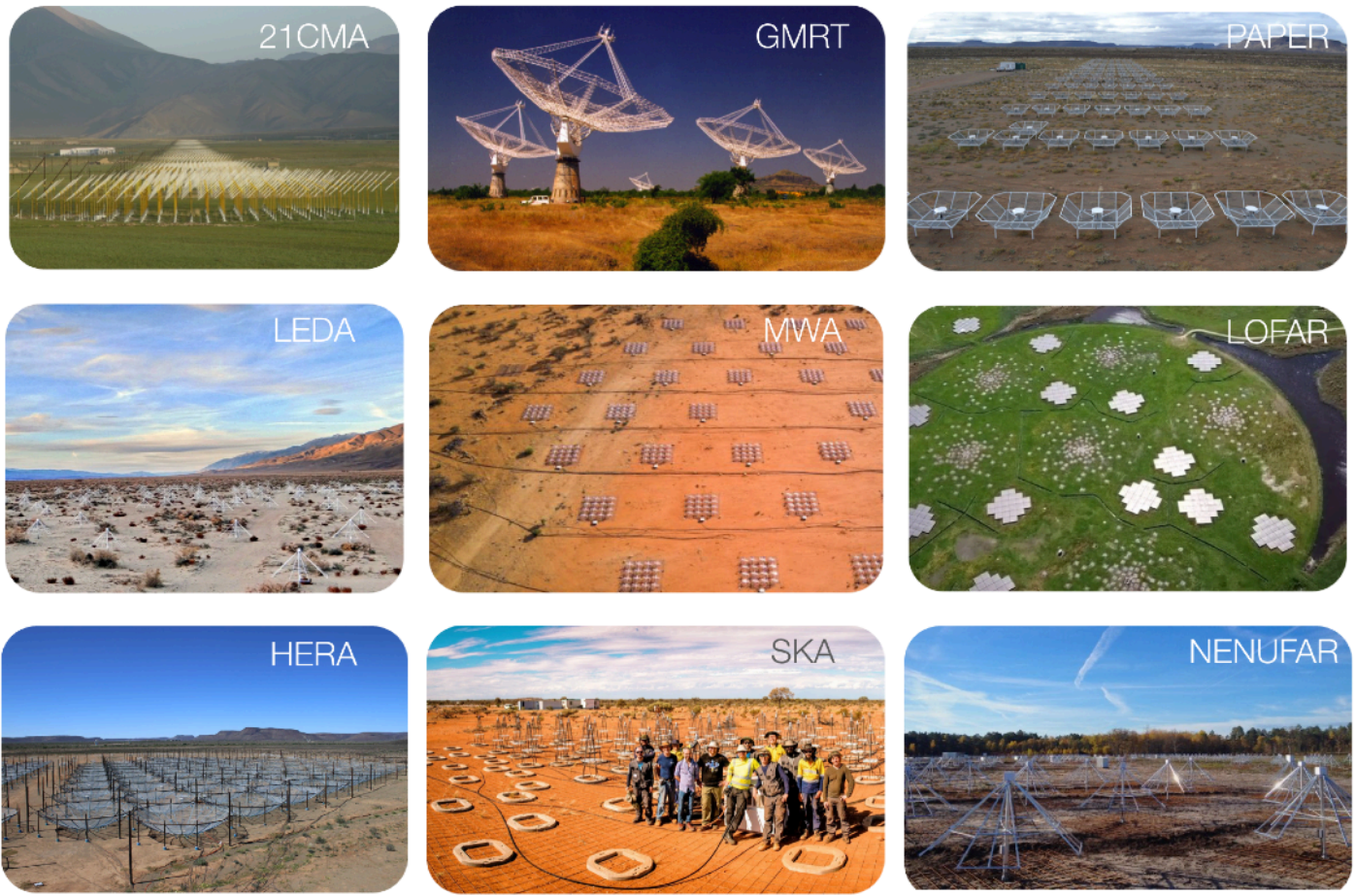

Fig.3: Panel of the current and planned 21-cm signal experiments (PAPER is decommissioned), focussing largely on probing the Epoch of Reionization (z~6-10) and late Cosmic Dawn (z<25), in no particular order.

**PAPER[4]:** The *Precision Array to Probe the Epoch of Reionization* was located in Karoo reserve area in South Africa. It was conceived to be a reconfigurable interferometric array to be deployed in stages of increasing number of elements (up to a maximum of 128 elements). It used individual 2 m dipoles sensible to the 100-200 MHz range that were first deployed in pseudo random configurations to test calibratability and observe properties of foreground sources (e.g., Parsons et al., 2010; Pober et al., 2013, Kohn et al., 2016). The PAPER team spearheaded a novel approach to the detection of the 21 cm signal by adopting highly redundant array configurations in order to maximise sensitivity on a number of power spectrum modes (Parsons et al., 2012). The drawback of redundant arrays is that they provide limited imaging capabilities and, therefore, limited foreground modelling capabilities. This led to the adoption of the so-called avoidance method: smooth-spectrum foregrounds are expected to be confined to a specific region of power spectrum space, leaving the remaining uncontaminated; this uncorrupted area is where the 21 cm signal detection should be attempted (e.g., Parsons et al. 2012). Power spectrum results were published and currently revisited in an upcoming publication (Kolopanis et al., 2019, in prep.).

**LOFAR[5]:** The *Low-Frequency Array* (LOFAR; van Haarlem et al. 2013) has the detection of the 21-cm signal of neutral hydrogen during the Epoch of Reionization (EoR; see, e.g. Patil et al. 2017) as one of its key-science drivers. LOFAR is also a pathfinder to the Square Kilometre Array (in particular SKA-low). The few-km inner core of LOFAR High Band Antenna (HBA; 110-240MHz) system consists of 48 stations each having 24 tiles of each four by four cross-dipoles. The large collecting area in the core and the high filling factor of LOFAR-HBA make it particularly suited for the detection of diffuse low-surface brightness emission of neutral hydrogen during the EoR. Besides

---

[4] http://eor.berkeley.edu/

[5] http://www.lofar.org/

LOFAR observing in standard station beam-forming mode, it can also cross-correlate all 576 HBA tiles or LBA dipoles (see below) of the inner 12 stations, yielding an extremely large field of view (e.g. Gehlot et al. 2019a) and hence probe a larger volume of the Universe, reducing sample variance that dominate the largest scales of the 21-cm signal. The Low-Band Antenna system (i.e. (10)-30MHz) could, in principle, detect the 21-cm signal from the Cosmic Dawn (CD) as well (e.g. Gehlot et al. 2019b), but only if the 21-cm signal exceeds nominal predictions of 10-100 mk$^2$ at k=0.1 cMpc$^{-1}$ by a significant fraction (e.g. Fialkov et al. 2018, 2019). The reason is the extremely bright Galactic foregrounds that cause a high noise level. The second reason is that ionospheric effects are far more significant at lower frequencies, causing signal distortions. Currently, LOFAR has yielded the deepest upper limits on the 21-cm signal from both the EoR (i.e. (79 mK)^2 for k=0.06 h Mpc$^{-1}$ at z~10; Patil et al. 2017) and the CD (z~25; Gehlot et al. 2019b). Observing the EDGES2 21-cm signal (see below) by cross-correlating all 576 dipoles of the inner 12 LBA stations (i.e. AARTFAAC; Gehlot et al. 2019a) is ongoing.

**MWA[6]:** The *Murchison Widefield Array* (Tingay et al. 2013, Wayth et al. 2018) is a low-frequency telescope in the Western Australian desert, on the same site of the future SKA. It comprises 256 tiles of 16 dual-polarisation dipoles in a regular 4x4 pattern, spread over 5 km. Detection of the signal from the EoR is one of its primary science goals, and it contains two subarrays of 36 tiles in a redundant hexagonal pattern to maximise sensitivity to EoR spatial scales, and provide additional instrument calibratability. The MWA is pursuing a statistical detection of the spatial fluctuation power spectrum, and has auxiliary programs to provide early results from cross-correlation studies, and pursuit of alternative statistics. Its large, 25 degree field-of-view improves sample variance, but at the cost of substantial leakage of power from sources close to the horizon into the EoR dataset. Smooth bandpass calibration, wide-field foregrounds and adequate sky models remain the primary challenges for undertaking EoR science with the MWA. Nonetheless, the EoR collaboration has published competitive upper limits (Beardsley et al. 2016) to the power spectrum on 10 arcmin to 1 degree scales at z = 6.5 – 9, and is also pursuing some analysis in the Cosmic Dawn at z = 15 – 18 (e.g. Ewall-Wice et al. 2016).

**OVRO-LWA[7]:** The *Owens Valley Radio Observatory Long-Wavelength Array (OVRO-LWA)* is located in Owens Valley, California. It operates in the 30-88 MHz frequency range corresponding to z~15-46, seeking to detect the 21-cm signal from the Cosmic Dawn. It aim to observe both the global signal (the experiment *LEDA, Large aperture Experiment to detect the Dark Ages*) via (five) individual dipoles equipped with custom-built calibration sources and 21-cm fluctuations via an array of 256 dipoles. Dipoles are pseudo randomly distributed to achieve an essentially filled array within a ~200m diameter core, providing excellent imaging capabilities to Galactic diffuse emission - the brightest foreground component. The OVRO-LWA approach to measure the 21 cm signal can be versatile, allowing to image and subtract foregrounds but also to isolate them in the power spectrum domain without any specific modelling. Current simulations shows that if IGM heating occurs efficiently at z ~ 16, OVRO-LWA would be able to detect the 21-cm power spectrum at k ~ 0.1 Mpc$^{-1}$ with a ~10 sigma signal-to-noise ratio in 3000 hours. First observations have set a $10^8$ (mK)$^2$ upper limits on the 21 cm power spectrum at k = 0.1 Mpc$^{-1}$ at z = 18.4 (Eastwood et al. 2019). Limits on the global signal can, and have, also been set with outrigger dipoles (e.g. Bernardi et al. 2015).

**NenuFar[8]:** *New Extension in Nançay Upgrading LOFAR (NenuFAR;* Zarka et al. 2015), situated in Nançay (France), is an extension of LOFAR but also a standalone instrument in the low-frequency range (10-85 MHz). Antennae were modelled on the LWA design whereas preamplifiers were designed in France. Antennas are distributed in 96 mini-arrays (>56 in place) each with 19 dual-polarised elements, densely covering a disk of 400 m in diameter, making it the most sensitive low-frequency array currently operations. Several mini-arrays are situated at distances of 2-3 km. Receivers will include the LOFAR backend, a local beam-former and a COBALT correlator. The NenuFAR concept has many points in common with GURT (the Giant Ukrainian Radio Telescope), with which it shares some technical studies, an its exploitation will benefit from a coordination with UTR-2. In the summer of 2019, a Key Science Programme has started (>1000h of observations in 2019-2020) with NenuFar to observe the 21-cm signal from the Cosmic Dawn, in particular to confirm or reject prediction made from the EDGES feature (Barkana 2018; Fialkov et al. 2018).

---

[6] http://www.mwatelescope.org/

[7] http://www.tauceti.caltech.edu/LWA/; see also for LEDA: http://www.tauceti.caltech.edu/leda/

[8] https://nenufar.obs-nancay.fr/en/homepage-en/

**HERA[9]:** The *Hydrogen Epoch of Reionization Array (HERA)* is an array currently under construction in the Karoo reserve area in South Africa. HERA is built following the approach used for PAPER: a highly redundant array to maximise the sensitivity on a number of power spectrum modes measured using the avoidance approach. In order to increase the sensitivity with respect to PAPER, it employs 14 m diameter dishes that, in the final configuration, will be densely packed in a highly redundant hexagonal array configuration of ~350 m diameter. HERA is built with the purpose to provide a complete statistical characterisation of cosmic reionization: its high brightness sensitivity configuration leads to a significant power spectrum detection in the Mpc$^{-1}$ range throughout Reionization (i.e., Pober et al., 2014; deBoer et al., 2017), fully constraining the evolution of the IGM neutral Hydrogen fraction. As the avoidance approach does not take advantage of foreground modelling, particular attention was paid to prevent the instrumental frequency response from corrupting intrinsically smooth foregrounds (Ewall-Wice et al., 2015; Patra et al., 2018). HERA is currently under construction, with more than 200 dishes deployed and science observations routinely carried out (Carilli et al. 2018, Kohn et al. 2018). New feeds that extend the sensitivity to the 50-250 MHz (i.e. enabling observations of the Cosmic Dawn) are currently deployed for testing. In summary, HERA is planned to deliver a complete characterisation of cosmic Reionization and to attempt the detection of the Cosmic Dawn. Given its redundant configuration, imaging capabilities remain limited and will be the target of a next generation experiment.

**SKA[10]:** The low-band system (SKA1-Low) of the *Square Kilometre Array (SKA)*, is a low-frequency aperture-array radio telescope system, made up of ~131,000 wide-band antennas, half of which will be situated in a high-filling factor dense core of ~0.2 km$^2$. It will operate from 50 MHz to 350 MHz and be located in Western Australia. SKA1-low is particular suitable for sensitive 21-cm signal observations of both the Epoch of Reionization, where direct imaging can be done down to the mK level on ~10 arc minute scales, and of the Cosmic Dawn, where it is expected to yield detailed 21-cm signal power-spectrum measurements up to redshifts z~25 (Mellema et al. 2013; Koopmans et al. 2015). SKA1-Low (as HERA), despite its sensitivity and collecting area, however, remains unable to observe beyond z~25 because of its frequency cutoff at 50MHz, motivated both by the presence of (strong) human-made RFI and very large errors induced in the data by the time-varying ionosphere.

> Whereas the above instruments aim to measure the spatial fluctuations of the 21-cm signal, the instruments below aim to measure it global sky-averaged brightness temperate (see above). We shortly discuss EDGES and SARAS. Besides these two extremely precisely designed and leading mono/dipole instruments (with deepest limits or claimed detection), a range of other global 21-cm signal experiments are currently ongoing such as e.g. PRIZM, BIGHORNS and SKIHI, which we will not further discuss because they ceased operation or not yet set competitive limits.

**EDGES[11]:** The U.S.-led Experiment to *Detect the Global EoR Signature (EDGES)* is located in Western Australia. Over the last decade, it has operated four instruments covering frequencies spanning the Cosmic Dawn and Reionization eras. Each instrument consists of a well-calibrated radio receiver connected to a single compact dipole-like antenna, chosen to minimise frequency-dependent effects. Building on techniques similar to those employed in other microwave measurements (Hu & Weinreb 2004, Belostotski 2011), EDGES has demonstrated end-to-end absolute measurement calibration (Rogers & Bowman 2012, Monsalve et al. 2017a) at the levels of accuracy needed for 21cm observations. The experiment has disfavoured rapid Reionization (Bowman et al. 2010, Monsalve et al. 2017b) and specific astrophysical models of early star formation, particularly those with little or late X-ray heating (Monsalve et al. 2018, 2019). In 2018, the EDGES team reported the first evidence for detection of the redshifted 21cm signal. They found an absorption profile in the sky-averaged radio spectrum centered at a frequency of 78 MHz, corresponding to redshift 17, and with an amplitude of 0.5K (Bowman et al. 2018). The amplitude of the observed profile is more than a factor of two greater than the largest standard predictions and suggests, surprisingly, that either the gas between early stars and galaxies was significantly colder than expected or that the background radiation temperature was hotter than expected. A number of possible explanations have been proposed, including interactions between baryons and various candidate dark matter particles (e.g. Barkana 2018, Munoz & Loeb 2018), as well as possible astronomical sources capable of producing an intense radio synchrotron background in the early Universe (e.g. Ewall-Wice et al. 2018). Efforts are presently underway in the 21-cm community to confirm the EDGES detection.

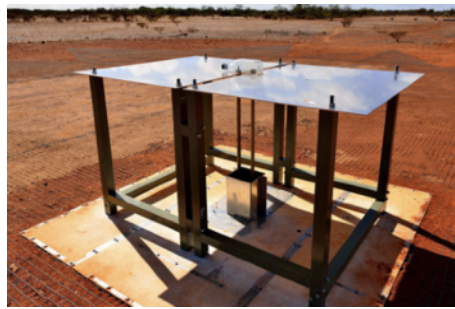
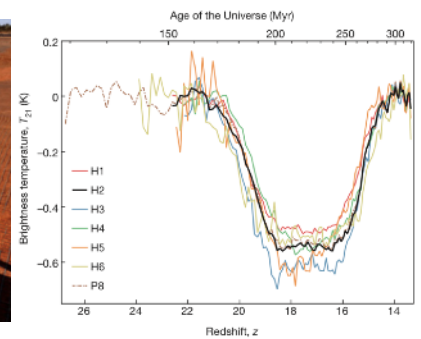

Fig.4: The EDGES instrument (right) and its claimed global 21-cm signal detection (right)

---

[9] https://reionization.org/

[10] https://www.skatelescope.org/   See also the SKA science book at https://www.skatelescope.org/books/

[11] http://loco.lab.asu.edu/edges/

**SARAS[12]:** At the Raman Research Institute in India, a progression of radiometers have been constructed and deployed using *Shaped Antennas to measure the background RAdio Spectrum (SARAS)*; collectively referred to as the SARAS radiometers. The radiometers have optimised their efficiencies in the 50-200 MHz band, with the goal of detecting global spectral distortions from redshifted 21-cm signals from Cosmic Dawn and Reionization. The SARAS antennas have all been electrically small thus providing frequency independent beams and avoiding mode coupling of sky spatial structures into confusing spectral structures. The shaping of the antenna elements was aimed at providing efficiencies that are maximally smooth (Satyanarayana Rao et al., 2017), so that the smooth foregrounds retain their smoothness in detected spectra without confusing any embedded 21-cm signal. The first SARAS experiment used a fat-dipole antenna (Patra et al., 2013). Recognising the advantages of monopole antennas over dipoles in their frequency independence, the improved SARAS 2 devised a shaped monopole antenna (Singh et al., 2018a). Deployed in the radio-quiet Timbaktu Collective in Southern India, data in the 110-200 MHz band was examined for signatures of cosmological Reionization. The class of cosmological models (from the atlas of Cohen et al., 2017) in which heating of primordial gas is inefficient---leading to deep absorption signals---together with rapid Reionization was rejected by the SARAS measurements (Singh et al., 2017; Singh et al., 2018b). The progress in design for avoidance of spurious structures from the relatively intense foreground, and precision calibration methods that avoid receiver systematics, paved the way for a ground-based SARAS3 that is optimised for the 50-100 MHz band. Additionally, recognising the maturity in the ground experiments, the Indian space agency – ISRO – has provided the SARAS team pre-project funding for development of a lunar orbiter mission – PRATUSH.

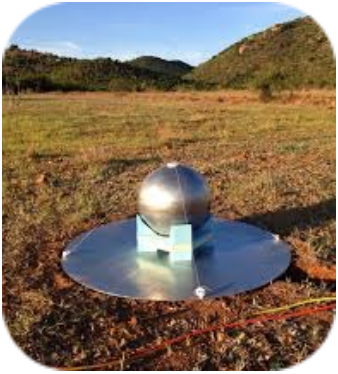

Fig.5: The SARAS instrument.

## 6. Current/Planned Space-based Observatories

**Motivation for space-based 21-cm signal observations:** Although current instruments (previous section) are slowly overcoming systematics related to the instrument (e.g. gain drifts, beam variations, etc.), the phase errors due to the ionosphere and still well beyond the ionospheric plasma frequency cutoff (~5-10MHz), the extremely bright (extra)Galactic foregrounds (hundred to thousands of Kelvin) that outshine the 21-cm signal (few to tens of mK), and man-made interference, each of these effects become increasingly stronger toward lower frequencies. Even in the deserts of the Karoo desert of South Africa and Western Australia, one can not escape RFI (e.g. from satellites, airplanes, etc.), and below ~50MHz the ionosphere becomes too turbulent to reach thermal-noise limited results. Even the slightest error of order $\sim 10^{-(4-6)}$ in the instrument or ionospheric corrections will extremely detrimental for any high-redshift 21-cm experiment. Each of these potential show-stoppers motivate a space based radio array.

✓ **No ionosphere:** Removing the detrimental impact of the ionosphere on the measurement electric field at each radio receiver, leading to both phase an amplitude errors, allows one to push observations to much higher redshifts, $z>>25$, potentially even below the ionospheric cutoff to the early stages of the Dark Ages ($z\sim200$, $\nu\sim7$ MHz). Although the lunar surface might be mildly ionised, this will not be the case in lunar orbit, in lunar L2, nor in outer space. On earth, the ionosphere requires second to minute time-scale calibration at ~100MHz, and even faster at lower frequencies, which will be infeasible for Dark Ages 21-cm observations. Moreover, on such short time-scales the instrumental signal-to-noise is not sufficient to solve for all unknowns of the ionosphere, effectively leading to speckle noise (e.g. Vedantham et al. 2014; Vedantham et al. 2016).

---

[12] http://www.rri.res.in/DISTORTION/index.html

- ✓ **Eliminate RFI:** Interference signals limit the frequencies observable from earth (some full excluded such as in the FM band, DABs, etc.; see e.g. Offringa et al. 2013, 2015). RFI can also occur as very low level, currently not seen in the data, and hence can be a potential showstopper for deep Dark Ages 21-cm signal observations, requiring dynamic ranges of ~$10^8$ in the data (80dB). In space, e.g. on the far-side of the moon, in lunar orbit (earth/sun shadow zone), or in the distant Earth-Sun L2 point, RFI from can either be largely eliminated.
- ✓ **Stable Instrument:** In space the environment of the instrument is shielded from weather, and temperature variations can be controlled better or happen slowly. This leads to a much more stable and controlled instrument environment, allowing for a slower calibration duty cycle (thus higher S/N), unlike on the ground where humidity and large (day-night) temperature variations continuously impact the performance of the instrument over the duration of the experiments which can be many years of observations. To correct for these errors, careful direction dependent and full-Stokes calibration is needed (e.g. Yatawatta et al. 2018).

Each of these have motivated the developed a a wide range of space-based 21-cm signal experiments, some focussing on the *global 21-cm signal* (sky-averaged; in line with e.g. EDGES and SARAS) via well-designed single receiver instruments (di/tripoles), and others focussed on the *spatial fluctuations of the 21-cm signal* (in line with e.g. LOFAR, MWA, etc.). Below we shortly summarise these ongoing efforts (some realised, some in planning phase):

**DAPPER[13]:** "Burns et al. (2019a) proposed a SmallSat low-frequency experiment called the Dark Ages Polarimeter PathfindER (DAPPER) to fly in conjunction with NASA's accelerated lunar exploration program. DAPPER is proposed to observe at frequencies 17-38 MHz (z~83-36). It will measure the amplitude of the 21-cm spectrum to a level that will distinguish between the standard cosmological model and models with additional cooling derived from current EDGES results. DAPPER's science instrument consists of dual orthogonal dipole antennas and a tone-injection spectrometer/polarimeter based on high heritage components from the Parker Solar Probe/FIELDS, THEMIS, and the Van Allen Probes. DAPPER will be deployed from the vicinity of NASA's Lunar Gateway or in cis-lunar space and descend to a 50×125 km lunar orbit using a deep-space spacecraft bus that has both high impulse and high delta-V. This orbit will facilitate the collection of 4615 hours of radio-quiet data over a 26.4 month lifetime. DAPPER will search for divergences from the standard model that will indicate new physics such as heating or cooling produced by dark matter. The Cosmic Dawn trough in the redshifted 21-cm spectrum is affected by the complex astrophysical history of the first luminous objects. Another trough is expected during the Dark Ages, prior to the formation of the first stars and thus determined entirely by cosmological phenomena (including dark matter). DAPPER will observe this pristine epoch (17-38 MHz; z~83-36), and will measure the amplitude of the 21-cm spectrum to the level required to distinguish the the standard cosmological model from that of additional cooling at >5-sigma. In addition to dark matter properties such as annihilation, decay, temperature, and interactions, the low-frequency background radiation level can significantly modify this trough. Hence, this observation constitutes a powerful, clean probe of exotic physics in the Dark Ages. A second objective for DAPPER will be to verify the recent EDGES results for Cosmic Dawn, in the uncontaminated environment above the lunar farside, with sparse frequency sampling from 55-107 MHz (z~25-12)."

**FareSide[14]:** "*FARSIDE (Farside Array for Radio Science Investigations of the Dark ages and Exoplanets*) is a Probe-class concept (Burns et al. 2019b) to place a low radio frequency interferometric array on the farside of the Moon. A NASA-funded design study, focused on the instrument, a deployment rover, the lander and base station, delivered an architecture broadly consistent with the requirements for a Probe mission (about $1.3 billion). This notional architecture consists of 128 dual polarisation antennas deployed across a 10 km area by a rover, and tethered to a base station for central processing, power and data transmission to the Lunar Gateway. FARSIDE would provide the capability to image the entire sky each minute in 1400 channels spanning frequencies from 100 kHz to 40 MHz, extending down two orders of magnitude below bands accessible to ground-based radio astronomy. The lunar farside can simultaneously provide isolation from terrestrial radio frequency interference, auroral kilometric radiation, and plasma noise from the solar wind. It is thus the only location within the inner solar system from which sky noise limited observations can be carried out at sub-MHz frequencies. This would enable near-continuous monitoring of the nearest stellar systems in the search for the radio signatures of coronal mass ejections and energetic particle events, and would also detect the magnetospheres for the nearest candidate habitable exoplanets. Simultaneously, FARSIDE would be used to characterise similar activity in our own solar system, from the Sun to the outer planets, including the

---

[13] https://www.colorado.edu/ness/dark-ages-polarimeter-pathfinder-dapper; Test largely taken and adapted from https://arxiv.org/pdf/1907.10853

[14] Text largely taken from https://arxiv.org/abs/1907.05407

hypothetical Planet Nine. Through precision calibration via an orbiting beacon, and exquisite foreground characterisation, FARSIDE would also measure the Dark Ages global 21-cm signal at redshifts z~50-100."

**DSL[15]:** In the *Discovering the Sky at the Longest Wavelengths (DSL)* mission concept (Chen et al. 2019), a constellation of micro-satellites circling the Moon on nearly-identical orbits, form a linear array while making interferometric observations of the sky. The mission can map the sky below 30 MHz, which is still largely unknown. Although the sensitivity of such an array is insufficient to detect the fluctuating 21cm signal from the dark age, it could make a useful first step by mapping out the foreground, and also probe the dark ages through a precision global spectrum measurement using single antenna. The observations will also be useful in a number of other fields, such as the study of Sun and planets, the interstellar medium, extragalactic radio sources, etc. In the DSL concept, a larger "mother" satellite leads or trails 5 to 8 smaller daughter satellites. The daughter satellites take the radio observation, and pass the data to the mother satellite using microwave link higher frequency bands. The microwave also serves for clock synchronisation and distance ranging. The mother handles the interferometry computation and communication with the Earth. The relative position of the daughters are determined from the ranging information and angular measurement of the star sensor cameras. A lunar orbit mission would be simpler and less expensive than a lunar surface mission of similar capacity: it does not need to land on the moon, thus it saves the required weight of the landing system. Furthermore, the low lunar orbit period is only a little more than two hours, there is no need deal with long lunar nights. and conventional solar power suffices. The data can be transmitted back to Earth during the time when the satellites are at the near side part of the orbit, there is no need to have relay satellites. The whole DSL constellation can be launched with a single CZ-2C rocket. The DSL project is now undergoing background prototype study.

**NCLE[16]:** On May 21st, 2018, a Chinese long March 3 rocket was launched from the Xichang launch base which brought the Queqiao relay satellite, as part from the Chang'e 4 mission, to space. The relay satellite is now one year behind the moon in the Earth-Moon second Lagrange point, and was used during the historical landing on the Lunar far-side in January 2019. However, the Queqiao satellite also caries a low-frequency radio instrument payload, the *Netherlands-China Low frequency Explorer (NCLE)*. NCLE is designed, built and tested by a Dutch consortium comprised of the Radboud University (PI Falcke, Dept, PI Klein Wolt), ASTRON (Co-PI Boonstra) and ISIS (Delft), in close collaboration with the National Astronomical Observatories of the Chinese Academy of Sciences (NAOC, Co-PI Ping). The instrument has three main components. The 3 antenna units are each 5 meter long carbon-fibre monopoles (Figure 6) that can be switched into dipole mode. The analogue system is sky noise limited in the 2-50 MHz regime, but the system is sensitive in the 80 kHz to 80 MHz regime, and in order to effectively deal with RFI and EMI peaks this range is split in 3 bands (16 channels 7.5-0.9 kHz): < 3MHz, 1-60 MHz and 60-80 MHz. There are six 1 MHz Hi-pass analogue modes and 10 MHz High-pass modes, and two 3 MHz low-pass analogue modes. The digital system has a 14-bit ADC (120 MHz), with full polarisation, 250 GB of on-board storage capacity and in total 39 digital modes. There in total 9 different science cases, ranging from Solar bursts, Earth RFI to Dark Ages and Cosmic Dawn science, that can be addressed by choosing the right combination of analogue and digital modes which will set the frequency regime of interest, the spectral and time resolution and the total observing time. While NCLE opens up a virtually unexplored frequency domain for a wealth of interesting science cases, the RFI characterisation is one of the most important technical objectives. NCLE will characterise the RFI and general noise environment behind the moon, a site that has been identified as being ideal for future low frequency radio missions. The major science potential will be the attempt to detect the global Dark Ages and Cosmic Dawn signals, regarded as the "holy grail" of cosmology, but NCLE will also provide unique measurements of bright sources like the Sun and Jupiter and attempt to create the most accurate (up to a few degrees) map of the sky at frequencies below ~10 MHz. While being in space already since May 2018, the NCLE instrument was only allowed to start the commissioning after the Chinese lander mission on the lunar far-side had progressed significantly. Currently, the Dutch NCLE team is analysing the first test data (Figure 7) from the instrument and has started working on the commissioning. The first data shows that the NCLE hardware is still in a perfect condition, and that on top of the broad band variations caused by the instrument response there are sharp noise peaks most likely related to the onboard electronics of the Queqiao satellite. Note that this data was taken during a one-minute integration end-to-end test that was concluded successfully, and that during that time the antenna units are not deployed. After careful calibration of the instrument response as a function of its orbit around the Earth, the antennas will be deployed in a step wise approach first to 50 cm and then in small increments to the full 5 meters. This is currently planned for October 2019 and will be followed by another full month of observations to characterise the instrument with antennas deployed for each position in its orbit. Calibration of the NCLE instrument is done by using both an internal calibration source and know astronomical sources, and is expected to be completed towards the end of 2019 after which the first science observations are planned.

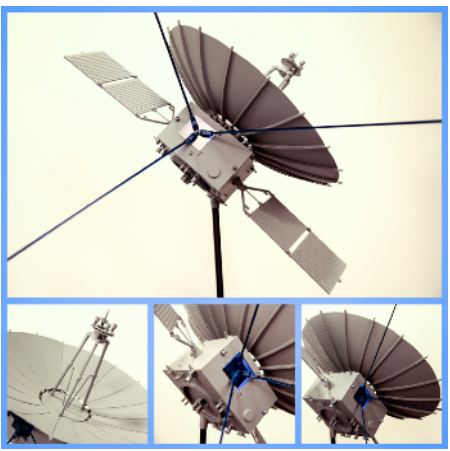

Fig.6: Rendered view of NCLE, the tripole attached to the Chang-e'4 relay satellite.

---

[15] http://www.issibj.ac.cn/Program/Forums/DiscoverSky/201808/t20180829_196751.html

[16] https://www.ru.nl/astrophysics/radboud-radio-lab/projects/netherlands-china-low-frequency-explorer-ncle/

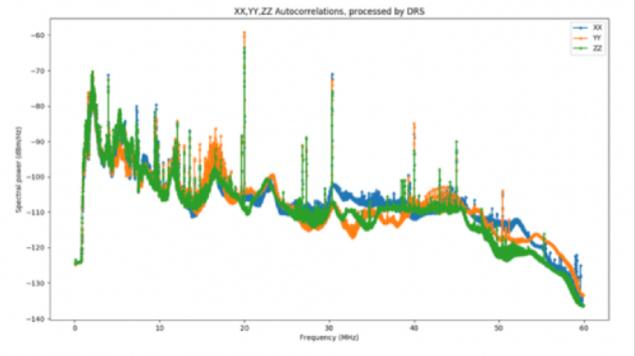

Fig.7: First data taken during an end-to-end test in January 2019. The different colours represent the autocorrelations from the 3 antenna units. On top of the broad band variations, with represent the variations in the response, there are clear sharp noise peaks most likely coming from electronic systems on the spacecraft.

## 7. CoDEX — Space/Lunar-based mission design considerations

Having discussed ongoing ground and space-based 21-cm instruments, *we now consider some generic mission design considerations that enable one to observe the 21-cm signal from the Dark Ages (z>25)*. These requirements are sufficiently stringent and inclusive that they also allow the lower-redshift 21-cm signal from the Cosmic Dawn (z<30) to be observed. To set the scene, but not to be complete in our analysis, we present *general requirements* for two representative redshifts (see Fig.1). We generically refer to such a mission concept as *CoDEX, the Cosmic-dawn Dark-ages EXplorer*. CoDEX should be able to

  A. measure the 21-cm signal power spectra at the end of the Dark Ages (z~30) with S/N>10 and image the 21-cm signal (tomography) with S/N > 10, over at least one dex in k-modes.
  B. measure the 21-cm signal power spectra at the peak of the Dark Ages (z~50) with S/N>10 and image the 21-cm signal (tomography) with S/N > 10, over at least one dex in k-modes.

Below we calculate the S/N ratios of CoDEX assuming a 5-year mission lifetime, BW=10MHz, and full-sky imaging.

**Basic Instrument Requirements of CoDEX**

The most basis technical requirement of CoDEX are tied to the extreme surface brightness sensitivity required to reach the faint (mK-level) 21-cm signal levels in the presence of very bright polarised foregrounds. This effectively translates into stringent requirements on the

  (1) collecting area ($A_{coll}$) —  the total integrated effective area inside the core of the array,
  (2) filling factor ($f=A_{coll}/A_{core}$), inside the core of the array, and
  (3) field of view $\Omega=\lambda^2/A_{eff}$, where $A_{eff}$ is the effective collecting are per receiver (e.g. dipole) and $\lambda$ the wavelength at which one observes.

We refer to Koopmans et al (2015) and Mellema et al (2013) for details on the error-budget of the 21-cm signal power spectrum scales with these three specifications. To calculate the S/N, we use the more detailed formalism of McQuinn et al. (2006) and assume the Galactic foregrounds also used for SKA models. The results are listed in Table.1.

This results shows that a S/N > 10 can be reached over a very wide range of scales and tomography can be performed as well for z=30 if CoDEX has a collecting area of at least 10 km², although some more modest targets (power-spectrum) can be reached at z=30 with a collecting area of 1 km². However, at least 100 km² collecting area is needed for tomography at z=50 with S/N>10 over k~0.01-0.1 Mpc$^{-1}$, which ultimate goal of CoDEX.

| CoDEX Mission | Dark Ages z=30, Power Spectra | Dark Ages z=30, Tomography | Dark Ages z=50, Power Spectra | Dark Ages z=50, Tomography |
|---|---|---|---|---|
| CoDEX (1 km²) M-class | S/N~10 for k~0.01-0.1 | S/N~5 for k=0.01 | S/N<1 | S/N<1 |
| CoDEX (10 km²) L-class | S/N~10-100 for k~0.01-1.0 | S/N~10-100 for k~0.01-0.1 | S/N>10 for k~0.01-1 | S/N>10 for k~0.01 |
| CoDEX (100 km²) L-class | S/N~100-1000 for k~0.01-1.0 | S/N~10-1000 for k~0.01-0.4 | S/N>100 for k~0.01-1 | S/N~10-100 for k~0.01-0.1 |

Table 1: CoDEX expected 21-cm signal S/N ratios for a 5 year mission lifetime assuming full-sky imaging. Image cubes have a depth of 10 MHz, centred on these redshifts. For the S/N calculation we assume the fiducial model as predicted from ΛCDM. The orange coloured boxes correspond to Fig.8. The green boxes reach our minimum power-spectrum requirements of S/N>10 over one dex in k-modes and the orange and blue boxes exceed these (either in S/N or k-mode range) for both power spectrum measurements and tomography.

In summary, we require:

- **A collecting area of ~10 km²** (f=1, 4) to reach a S/N>10 at k=0.01-0.1 Mpc$^{-1}$ **at z=30** within a 5-year mission lifetime, with 10MHz BW (i.e. redshift range), assuming all-sky imaging.

- **A collecting area of ~100 km²** to reach a S/N>10 at k~0.01-0.1 Mpc$^{-1}$ at **z=50** within a 5-year mission lifetime, with 10MHz BW (i.e. redshift range), assuming all-sky imaging.

*We conclude that to measure the 21-cm signal from the Dark Ages from z~30 up to z~50, any space-based low-frequency interferometer should have **a lifespan of at least 5 years, a collecting area of >1 km² (M-mission scale) for first results, expandable to 100 km² (L-mission scale), a filling factor of unity and all-sky. field of view** to reach the minimum S/N>10 for both power-spectrum and tomography over at least one dex. in spatial scale. The technical requirement of such a generic instrument (CoDEX) are summarised in Table.2.*

## Mission concept considerations

We now look in particular at a swarm of (cube)sats and/or connected receivers, either in sun-earth or lunar L2, in lunar orbit or on its surface. Each option has commonalities, requiring large collecting areas, a high filling factor and a large field of view. We again refer to this generic concept as *CoDEX, the Cosmic-dawn Dark-ages EXplorer.*

**Free-space mission concept:**

Based on previous designs for space-based, low-frequency, interferometry missions such as DARIS (Daris, 2010), SURO, HEIMDAL, DARE (Burns et al. 2012) and OLFAR (Engelen 2010), and most recently DEX (Klein Wolt et al. 2013), a generic space-based mission concept could be as follows.

- The individual antennas are (semi)omni-directional di or tripoles (active antennas), with their sensitivity optimised in the 1-80 MHz regime over a bandwidth of 40 MHz. An array of core antennae can be placed on inflatable/foldable space structures or connectable nano-satellites, putting them in a (regular) grid ('core') to provide a high filing factor (f~1) optimally sensitive to the very 21-cm faint signals, surrounded by outer antennae (e.g. free-floating satellites) to enable the higher resolution imaging.
- The free-floating antennas are mounted on nano or small satellites that provide power (solar panels) and basic processing (RFI excision, FFT and averaging algorithms) and communication. Together with a mother-ship, possibly mounted on or closely located to the core, they provide data storage, processing and communication with Earth. The constellation acts as a "swarm".

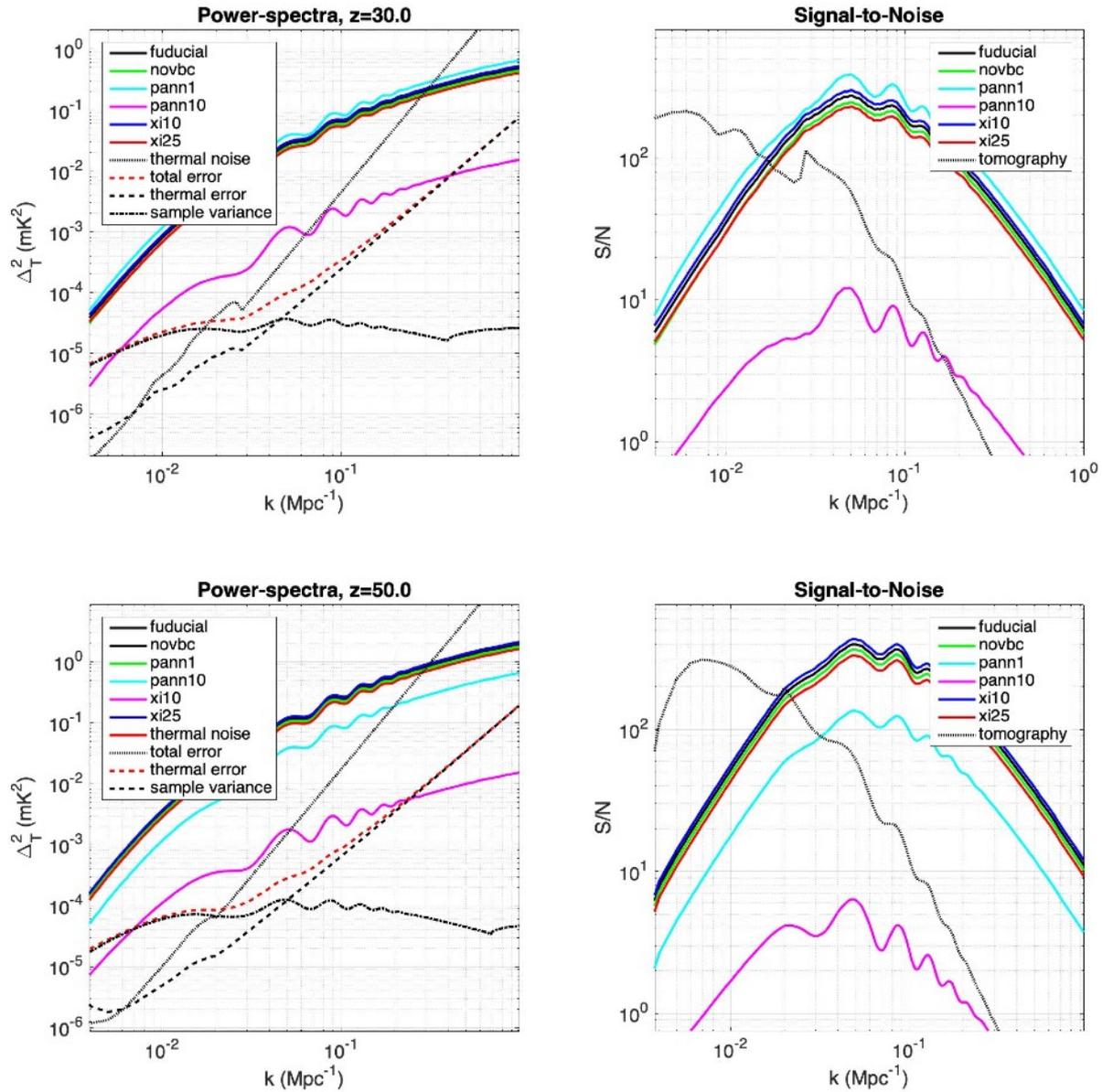

Fig 8. Top Left: Simulated power spectra of the 21-cm line based on Ali-Haïmoud, et al. (2014); Chen et al. (2016) (top solid solid lines; the fiducial model is in black (the top four power spectra are offset by 10% for clarity) and the expected sensitivity of the DEX array (red dashed line) for z=30 for a 10 km$^2$ core and a 5-year mission, using a BW=10MHz. Top Right: Signal-to-noise achieved. Bottom Left/Right: Idem for z=50 with a 100 km$^2$ core. In both cases, the dotted line indicated the thermal-noise power-spectrum. In case the 21-cm signal exceeds the noise power spectrum, direct imaging of those scales (tomography) can be done. We note that direct imaging on the largest scales is not affected by sample variance and can typically be done for scales larger than half a degree (k~0.1).

- The core provides the sensitivity for CD/DA observations, providing perfect uv coverage. The outer free-floating satellites/antenna provide superb imaging quality at higher angular resolutions. Due to the motion of the individual antennas, the baselines are constantly changing and complete coverage of the *(u,v,w)* plane is achieved.
- The array should be placed in a RFI-quiet location, for instance at the Sun-Earth L2 point or in an Earth-leading or Earth-trailing orbit, or in lunar orbit (similar to DLS). This will minimise the need for RFI-mitigation techniques.
- Each individual antenna should be calibrated carefully and the noise pattern (EMC) should be characterised in detail. Stringent satellite EMC measures need to be implemented, but previous single element radio missions have shown that this is feasible.
- Data processing can be done at a central mother-ship which has dedicated data processing facilities, more power available (larger solar panels) and which has a large data rate available for data transfer. The individual antennae then have to transfer the data to the mothership and have to perform on-board processing to reduce the data volume; for the core receivers the data-transport will be easier if mounted near the central processor.

- Alternatively, data processing and transfer can be arranged by the array of antennas, i.e. they act as a swarm, performing the signal processing in a distributed fashion. This has the advantage that with the increase of the array the processing power is increased, but also requires each individual element to be provided with significant processing and communication capabilities;
- For the communication between core antennas and the mother-ship, RF links can be used, for the individual antennas and the mothership, optical communication can be used;

| Technical Requirements | | Value / range | Comments |
|---|---|---|---|
| **TR01** | Frequency range | 1 – 100 MHz | Covers DA/CD |
| **TR02** | Frequency resolution | 1 kHz (RFI/burst mode) 10-100kHz (DA/EoR science) | RFI excision/Limit BW smearing/ Ancillary. Science |
| | Bandwidth | 50MHz (between 1-100MHz) tuneable | Covering z=50-20 in one go and provide calibration BW |
| **TR03** | Time resolution | 50 ms (RFI/burst mode) 10s (DA/CD science) | RFI excision/DA & CD science Burst mode/ Anc. Science |
| **TR04** | Core Area of the array | $A_{core}$ = 1-10-100 km$^2$ | With filling factor ~1 |
| **TR05** | Array elements | $3\times 10^{4-5-6}$ | Depends on optimal λ and $A_{core}$; most are in a "fully-filled" core |
| **TR06** | Number of baselines | $10^{9-11-13}$ | Possibly FFT correlation core |
| **TR07** | Baseline lengths | <=10 km (core) <=100km (outer) | "Satellite" outer stations up to 100 km. Core resolution <10' and <0.1' on long baselines, at at z=50 (better a lower z). |
| **TR08** | Element field of view | 4π sr with side-lobes for space array, about 2π sr for lunar array | "all-sky" coverage for a space array, one hemisphere per snapshot for a lunar array |

Table 2: CoDEX technical requirements

**Lunar-based mission concept:**

The CoDEX lunar concept design on the one hand draws from the previous proposed lunar mission concepts, but on the other hand will be based on the experience and expertise gained with Earth-based low-frequency interferometers such as LOFAR (van Haarlem 2013) and in the future SKA and HERA. In particular, the experience in the technical realisation of a large collecting area radio array in (often) remote locations, calibration of the instruments, RFI mitigation techniques and the handling and processing of large data volumes is essential for the development of CoDEX. In short, the CoDEX lunar interferometer design has the following characteristics:

- Interferometer array, consisting of about $3\times 10^{4-5-6}$ individual antenna elements together ultimately realising a staged 1-10-100 km$^2$ collecting area on the lunar surface, sensitive in the 1-100 MHz frequency regime.
- The individual antenna elements can consist of traditional dipole or tripole antennas (e.g. LRX, see Klein Wolt et al. 2012) or be placed on thin metal sheets that are rolled out on the surface (ROLSS, see Lazio et al. 2011).
- The location should be chosen carefully in order to provide temperature and gain stability as well as attenuation of RFI signals from Earth and Solar activity, preferred is the lunar far-side but South- or North Pole locations using shielding from mountains is also an option. Note that while the lunar far-side is preferred, it does require an additional orbiter for communication to the Earth.
- At a Lunar far-side location (similar to FareSide; see Section 6) the data processing and communication should be done from an orbiter or relay satellite (as for Chang-e'4/NCLE): it provides more power from solar panels and has better communications with Earth available compared to stationary platforms on the lunar far-side.
- Each individual antenna should be calibrated carefully and the noise pattern (EMC) should be characterised in detail.

# Key Technologies — some general considerations

Here we provide an overview of some of the key technologies that are required for the mission concept, and provide some thoughts on technologies that could help build a large interferometer in space or on the lunar surface:

**Nano-satellites & Swarm technologies**

The main design considerations for a free-floating astronomical low-frequency array in space relate to the physical characteristics of the interplanetary and interstellar medium. The system must consist of a high filling factor "core" plus a swarm of identical satellites spread over kilometric distances that will orbit faraway from terrestrial radio frequency interference. To reduce weight one needs to miniaturise the electronics and perhaps use nano satellites with masses between 1-10 kg. One approach is to use a swarm of satellites to establish a virtual telescope. In recent studies, such as DARIS (Saks et al., 2010) and FIRST (Bergman et al., 2009) it is shown that with extrapolation of current signal processing and satellite technologies, a low frequency radio telescope in space could be feasible. DARIS has shown that a 9-satellite cluster, with a centralised system can be implemented in lunar orbit with today's technology (see also DSL above). Similar to the OLFAR project (Bentum et al., 2009; Budianu 2015, Rajan 2016, Engelen 2016), CoDEX needs scalable autonomous satellite flight units. To achieve sufficient spatial resolution, the minimum distances between the satellites must be more than 10 km and due to interstellar scattering this maximum baseline is limited to 100 km, giving a resolution of 1 arc minute at 10 MHz. Each individual satellite will consist of deployable (e.g. tripole) antennas. As the satellites will be far away from Earth, communication to and from Earth will require diversity communication schemes, using all the individual satellites together.

**Inflatable space structures**

CoDEX requires a significant collecting area (staged in steps, e.g. 1-10-100 $km^2$) and while radio antenna and receiver technologies are well developed (TRL levels of 6 and higher), bringing them into space or to the moon is a costly and technologically challenging endeavour. In order to reduce the weight, the mechanisms required for the deployment and hence the costs of such a mission, the use of inflatable space structures is suggested. Another reason for using inflatable structures is the required filling factor in the core of the instrument. A swarm of antennas cannot easily obtain a filling factor of 1 without colliding and an inflatable structure can help to create the 1–10-100 $km^2$ with f=1 to starting doing interesting observations of the DA/CD. The first two missions using large inflatable structures launched by NASA in 1960, Echo 1 and 2, were successfully used as communication reflectors. Interest in inflatable space structures was renewed which the successful deployment of the NASA Inflatable Antenna Experiment (IAE) from space shuttle mission STS-77 in 1996 (Freeland et al., 1997). In addition, several mission concepts were proposed, for instance MIT and JPL's mission to develop an inflatable antenna structure for cube sats (Babuscia et al, 2014). For CoDEX the inflatable space structure technology can be adapted for a platform or deployment of the antenna system, comparable to the IAE concept. For instance, the scalable 1-10-100 $km^2$ required for CoDEX can be obtained by forming an array of Echo2-like antennas, or place them on flat foldable surfaces (e.g. "solar sail" material) surrounded by an inflatable structure to keep the area rigid. The latter would also allow a staged to increase collecting area. Given the current development in the commercial heavy-launch space flight, the deployment and realisation of CoDEX is becoming feasible in the near future from a technological and financial point-of-view.

**Additive manufacturing**

In order to minimise the number of launches and the payload required for construction of the radio-antenna array on the Moon or in space, the utilisation of resources available on the Moon is highly desirable. It can dramatically reduce the overall project cost, provided that the resources present on the lunar surface are adequate and that the technology is available to transform these resources into the desired products and structures. The lunar surface consists of a mixture of dust, soil and broken rocks commonly referred to as regolith. The regolith is composed of various oxide-containing minerals. Previous lunar missions have shown that the regolith composition varies in different regions of the Moon. The compositions in various regions can be found in (Basu 1998, Crawford 2015). The oxides of interest for in-situ construction of electrically conductive antennas are the metal-bearing ones (e.g. $TiO_2$, $Al_2O_3$, FeO). In the perspective of in-situ resource utilisation for construction of the antenna array, Additive Manufacturing (AM) techniques can be of high interest, as they allow efficient material use and can be automated. The European Space Agency has developed and is currently investigating AM processes for the construction of hardware and infrastructure, to support the establishment of human settlements on the Moon and Mars. Such processes could potentially be applied – after required maturation and validation in a relevant environment – to build supporting infrastructure for the antenna array installations. Ongoing and planned activities at ESA also aim at studying the possibilities of using and recycling materials brought from Earth for the mission, in particular polymers (ESA 2016). AM of polymers has been demonstrated on board the International Space Station, for relatively small parts. However, the use of such polymers to build the antenna arrays would require to make these materials electrically conductive (e.g. by addition of conductive particles). In-orbit AM of such conductive composite materials has not been demonstrated yet. In addition, using polymer resources brought from Earth for the construction of the radio-telescope would still require large payload mass. In order to build metallic antenna arrays using in-situ resources, the metallic elements would need to be extracted from the regolith. These metals could then be used as feedstock for AM processes. A range of AM processes exists for metallic products, using metallic feedstock in a powder or wire form. The adaption and validation of such processes for the lunar environment would need to be performed. Several techniques have been proposed for the extraction of metal resources from the regolith. These techniques are usually primarily intended to extract volatiles (oxygen, hydrogen or water) to support manned missions. But they often yield metals and other solids as by-products. The as-extracted metallic by-products could be used, provided that they have the required electrical conductivity for the antenna application. Alternatively, the by-products would need to be beneficiated to the required purity. The processes that would turn the extracted metals into antenna arrays also need to be developed for the lunar environment. In addition, given the large collecting area required for the antenna array (up to ~100 km$^2$) the means of mining, sieving and transporting the regolith to the desired manufacturing location need to be studied. Alternatively, a mobile AM system (e.g. mounted on a rover) needs to be developed. Energy transportation (e.g. power cables) from the energy harvesting sites to the manufacturing sites may also be required. Regarding the option of a space-based antenna array, no experience on in-orbit manufacturing of metallic products has been reported.

| Technologies | Current TRL | Expected TRL 2020-2030 | Expected critical developments |
|---|---|---|---|
| Radio Antenna (lightweight, foldable, inflatable) | 5 | 6-7 | Small, foldable light-weight structures are being designed in the OLFAR and ROLSS project that should fit a nano-satellite. Current TRL increasing activity is the NCLE antenna + deployment system that is being designed for the Chang'e 4 mission (launched in 2018, PI Falcke) |
| Radio Receiver (low-power, high processing, 200 MHz receivers) | 5-6 | 6-7 | Prototype radio receivers are expected to be tested in rocket flights in the near future, and similar systems will be tested in space environments (e.g. ISS). Further heritage is gained from ground-based low-frequency instruments such as LOFAR, SKA, MWA, LWA and space-based instruments such as LRO. Current TRL increasing activity is the NCLE receiver system that is being designed for the Chang'e 4 mission (launched in 2018, PI Falcke) |
| Digital processing system | 4-5 | 6-7 | Development of power-saving and smart algorithms to processes large quantities of data with significantly less power are currently ongoing in many Big-Data Science projects (CERN, ITER, LOFAR, SKA) |
| Optical communication | 6-8 | 7-9 | Optical communication and nano-photonics are expected to be employed in space industry (telecom) and has been tested since the 1970s (e.g. SILEX on ESA Artemis) |
| Swarm Technologies | 3-4 | 5-6 | OLFAR, inter-satellite communication, satellite control |
| Thin-film solar panels | 4-5 | 7-8 | Currently thin-film solar panels are considered for future missions with expected launch dates before L2 and L3 |
| Radio Frequency interferometry | 7-9 | 7-9 | Based on space-ground interferometers (HALCA and RadioAstron), and there's been some crude time-difference-of-arrival measurements (interferometric-like) using the THEMIS or Cluster spacecraft. |
| Inflatable space structures | 7-8 | 8-9 | NASA Echo1 and 2 missions, NASA inflatable antenna experiment (IAE, 1996) |
| Additive manufacturing | 4-5 | 7-8 | Energy requirement for the manufacturing process, extracting process (mining) from regolith, production of conductive materials, impact of vacuum, reduced gravity, the presence of abrasive and electrically charged regolith particles, as well as the influence of temperature variations on the lunar surface on the sintering process |

Table 3: TRL level for some of the CoDEX technologies.

**Scalability**

The realisation of a staged 1-10-100 km$^2$ CoDEX interferometer in space or on the moon is a significant technological challenge that should be approached in a step-wise fashion and depends on future developments in the areas of space transportation, light-weight inflatable antenna space structures, data communication and low-power-high-performance data processing and swarm technologies. The CoDEX concepts (both the space-based and the lunar-based concepts) are based on the same science case and have commonalities with respect to the data processing and operational modes, but also are equally flexible and scalable. In both cases individual elements can be added to the array which would increase the sensitivity and science output, and in some cases also the total data processing capabilities. Starting a 0.1 km$^2$ CoDEX array, as a pathfinder mission, would also immediately provide valuable science output; not only would the CoDEX path-finder open up the last unexplored frequency domain, but it would allow for the detection of the global Dark Ages signal as well address many of the secondary science cases (planetary radio emission, all sky survey etc). As elements are added to the CoDEX array, more science-domains will open.

**TRL Levels**

In Table 3 we present an overview of the technologies introduced in the previous section that are required for a CoDEX mission. In this table we have indicated the current TRL level, as well as the level expected to be reached at the end of the 2020-2030 timeframe, and the critical developments that are required to be able to reach that level. The TRL level are indications and are based on our observations and some discussions with experts. However, it is important to realise that some of the critical developments are happening now already, independently of the CoDEX initiative. We believe that CoDEX can benefit from these developments, but moreover could provide a very interesting (additional) science case to support them

## 8. Short Summary & Conclusions

In this White Paper, we have argued for the development of space-based technology that enables the detection of the 21-cm signal from neutral hydrogen at redshift up to z~50, during the peak of the Dark Ages. This will be impossible from the ground, where the ionosphere is prohibitively volatile to enable such observations. Cosmology with the 21-cm signal of neutral hydrogen at hight redshift has proven extremely difficult — as demonstrated by many of the current ground-based instruments aiming for a first detection of the 21-cm signal from the Epoch of Reionization and late Cosmic Dawn— but it promises an entirely new avenue to study the infant Universe (z>6), probing astrophysics of the first stars, galaxies, black holes and intergalactic medium during the Cosmic Dawn (z>10), and potentially unveiling entirely new physics during the Dark Ages (z>30). The reason for these difficulties has been the unstable environment on earth (weather, ionosphere, human-made RFI, etc) in which these ground-based instrument have to operate, the complexity of the instrument itself and the extremely bright partly polarised (extra)Galactic foreground that are up to 10$^6$ times brighter than the desired 21-cm signal.

*The Dark Ages, however, ranging in redshift from z~30 to ~200, is the absolute holy grail of 21-cm Cosmology.* During this era, the Universe was pristine and linear perturbation theory holds. Currently well-understood physics should therefore be able to predict the 21-cm signal from the Dark Ages with great precision, *except* if new physics manifests itself at the densities and scales probed at these high redshift. Deviations from these standard-model predictions are direct evidence for physics beyond the standard model. Furthermore, the information contained in the 21-cm signal contains many orders of magnitude more modes than incorporated in the CMB, which has already provided cosmologists with a wealth of new discoveries and led to two Nobel Prizes, or in future large-scale galaxy surveys such as those to be performed with LSST or EUCLID.

Uncovering the 21-cm signal during the Dark Ages will arguably be one of the most daunting and difficult experiments ever undertaken, on par with the recent discoveries of gravitational waves and the Higgs boson. Developing this new frontier is well worth the effort and could uncover fundamental new insights into the origin of the Universe, as well lead to exploration of new physics.